\documentclass[conference]{IEEEtran}
\IEEEoverridecommandlockouts

\usepackage{tabularx}
\usepackage{tikz}
\usepackage{pgfplots}

\pgfplotsset{compat=newest} 
\usepackage{algorithm}
\usepackage{algorithmic}
\usepackage{graphicx}
\usepackage{diagbox}
\usepackage{adjustbox}
\usepackage{balance}
\usepackage{csquotes}
\usepackage{framed}
\usepackage{multirow}
\usepackage{url}
\usepackage{amsmath}
\usepackage[table,xcdraw]{xcolor}
\usepackage{pgfplots}
\usepackage[many,listings]{tcolorbox}
\tcbuselibrary{breakable}
\usepackage{listings}
\usepackage{caption}
\usepackage{booktabs}
\usepackage{hyperref}

\pgfplotsset{compat=1.17}
\definecolor{myblue}{rgb}{0.4, 0.6, 0.8}
\definecolor{myyellow}{rgb}{1.0, 0.9, 0.6}
\definecolor{myorange}{rgb}{1.0, 0.7, 0.5}
\definecolor{mygreen}{rgb}{0.6, 0.8, 0.6}
\definecolor{tablegray}{HTML}{F6F6FA}

\definecolor{bg}{rgb}{1,1,1}
\definecolor{keyword}{rgb}{0.75,0,0}
\definecolor{string}{rgb}{0.58,0,0.82}
\definecolor{redtext}{rgb}{0.75,0,0}
\definecolor{numberbg}{rgb}{0.8,0.8,0.84}

\usepackage[normalem]{ulem}

\tcbuselibrary{skins}
\tcbset{RQsboxstyle/.style={
    width=\columnwidth,       
    colframe=gray,            
    colback=gray!10,          
    coltitle=black,           
    boxrule=0.1mm,            
    arc=2mm,                  
    auto outer arc,           
    boxsep=4mm,               
    drop shadow={black!50!white},  
    leftrule=1.5mm
}}

\tcbset{blueboxstyle/.style={
    width=1\columnwidth,       
    colframe=blue!45!black,        
    colback=blue!5,        
    coltitle=black,          
    boxrule=0.1mm,           
    arc=2mm,                  
    auto outer arc,           
    boxsep=4mm,   
    leftrule=1.5mm,
    drop shadow={black!50!white} 
        
}}

\tcbset{yellowboxstyle/.style={
    width=\columnwidth,       
    colframe=yellow!40!black,          
    colback=yellow!10,        
    coltitle=black,           
    boxrule=0.1mm,            
    arc=1mm,                  
    auto outer arc,           
    boxsep=1mm,               
    drop shadow={black!50!white} 
}}

\usepackage{tikz}
\usetikzlibrary{backgrounds}
\usepackage{graphicx}

\definecolor{myexact}{HTML}{C3B1E1}  
\definecolor{mynear}{HTML}{FFB347}   
\definecolor{myunique}{HTML}{B2D8B2} 

\definecolor{trainbg}{gray}{0.95}    
\definecolor{testbg}{gray}{0.90}     

\begin{document}



\title{From Data Leak to Secret Misses: The Impact of Data Leakage on Secret Detection Models}

\author{
    \IEEEauthorblockN{Farnaz Soltaniani}
    \IEEEauthorblockA{Technische Universität Clausthal,
    Germany\\
    farnaz.soltaniani@tu-clausthal.de}
    \and
    \IEEEauthorblockN{Mohammad Ghafari}
    \IEEEauthorblockA{Technische Universität Clausthal, Germany\\
    mohammad.ghafari@tu-clausthal.de}
}

\maketitle

\begin{abstract}

Machine learning models are increasingly used for software security tasks. These models are commonly trained and evaluated on large Internet-derived datasets, which often contain duplicated or highly similar samples. When such samples are split across training and test sets, data leakage may occur, allowing models to memorize patterns instead of learning to generalize. 
We investigate duplication in a widely used benchmark dataset of hard coded secrets and show how data leakage can substantially inflate the reported performance of AI-based secret detectors, resulting in a misleading picture of their real-world effectiveness.
\end{abstract}


\section{INTRODUCTION}\label{sec:introduction}

Machine learning models are becoming increasingly popular for addressing problems in software security~\cite{Soltaniani2026LLMforSBR,10.1145/3661167.3661263,10.1145/3727582.3728686, 11052790, FIROUZI2026112650}.
Many of these models are built and tested using datasets sourced from the Internet. However, large-scale sources, such as GitHub, may contain duplicate or highly similar samples~\cite{lopes2017dejavu} that, when divided into training and testing subsets for model development, can inadvertently overlap. This phenomenon, known as data leakage, enables models to memorize recurring patterns rather than generalize, leading to inflated performance metrics~\cite{277204,10759822, evertz-26-chasing}. 
%

We focus on a specific application of ML for security, namely the detection of hardcoded secrets. Such practices are prevalent in software systems~\cite{GitGuardian2025} and expose confidential resources to anyone who can access the codebase~\cite{Meli2019HowBC}.
Hence, early detection of secret leaks is essential to minimize the opportunity for attackers and reduce the chance of a breach.
To address this issue, researchers have developed AI models that detect hardcoded secrets~\cite{9794113, 11323433,Soltaniani2026HardcodedSecrets}.

Existing models for secret detection rely heavily on SecretBench~\cite{secretbench}, a large benchmark dataset of secrets, mined from GitHub repositories. The dependence on GitHub as a data source raises concerns about code duplication within the dataset, which may compromise the validity of secret detection models.
%
We study the extent of data duplication in SecretBench and assess its impact on the performance of secret detection models. 
Our analysis focuses on recent studies that identify secrets by leveraging both the secret value and the surrounding code context.
To understand how duplication affects different modeling paradigms, we build multiple models under varying levels of data contamination. In particular, we investigate the following research question:

\textbf{RQ:} To what extent does code duplication impact the performance and generalization of machine learning models for secret detection?

We identify exact and semantically similar duplicate code samples obtained from the SecretBench dataset.
We study both traditional models (e.g., Random Forest) and newer learning-based models, including deep architectures such as Long Short-Term Memory networks and Transformer-based models like GraphCodeBERT.
We examine the models under varying data contamination scenarios, including training with and without duplicated samples. 

We found that duplication is prevalent in SecretBench and that data leakage is not a minor issue; it distorts MCC scores across all models, though the effect varies by model family. The Random Forest model is most affected; it drops from 0.89 to 0.77 when exact duplicates are removed and further to 0.65 when both exact and semantically similar samples are excluded. 
Similarly, the LSTM model exhibits significant vulnerability to leakage, with its MCC decreasing from 0.92 on the original dataset to 0.77 once all leaky samples are removed.
GraphCodeBERT appears more robust to duplications when the model is fine-tuned fully, exhibiting only a 7\% decrease in MCC after duplicate removal.
%


%


In summary, this study shows that secret-detection models built and tested using SecretBench may appear more capable than they actually are, as widespread duplication in the dataset significantly inflates evaluation scores.
Therefore, practitioners and researchers must place greater emphasis on dataset hygiene, adopt evaluation protocols that expose robustness issues, and design models that can handle novel or modified secret patterns.

We release a replication package containing all scripts and results, along with the corresponding SecretBench record IDs.\footnote{\url{https://zenodo.org/records/18164088}} For ethical reasons, access to SecretBench must be obtained before the full pipeline can be replicated.

The remainder of this paper is organized as follows. Section \ref{sec:problemstatement} defines the problem statement and Section \ref{sec:Methodology} presents our methodology. We then detail the experimental results in Section \ref{sec:EXPERIMENTRESULTS} and discuss potential threats to validity in Section \ref{sec:ThreatsToValidity}. Finally, Section \ref{sec:Conclusion} concludes the paper.



\section{Problem Statement}
\label{sec:problemstatement}

Source code contains rich contextual information that models can exploit to identify secrets and reduce false positives produced by rule-based detectors. Recent work has therefore focused on building secret-detection models that leverage \emph{secret context}, namely the secret value together with its surrounding code~\cite{SourceCodeSecretsML,9794113,AssetHarvester,11323433}. However, duplication within these secret contexts can distort model performance and undermine reliability in real-world settings.
In this section, we analyze the extent of duplication in SecretBench, a widely used benchmark dataset for secret detection.



\subsection{Secret Context}
We define secret context as the 200 characters preceding and following each secret value. We selected this window size based on a recent automated secret detection study~\cite{11323433}.

\begin{figure}[htbp]
    \centering
\includegraphics[width=0.9\columnwidth, height=5.5cm]{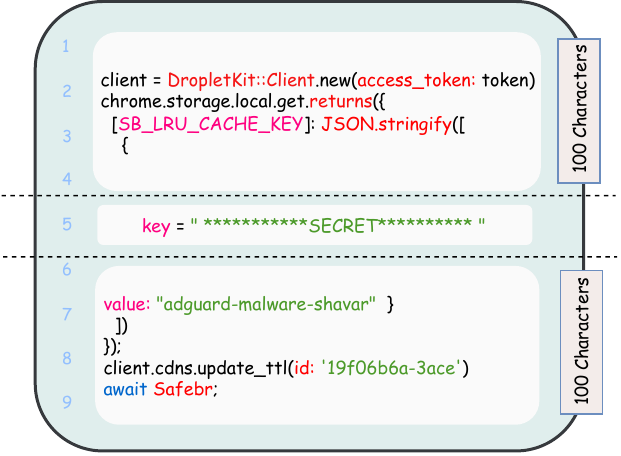}
    \caption{Example of a secret code context.}
    \label{fig:CodeContext}
\end{figure}

We exclude the secret string from this 200-character window to keep the context size consistent and to reduce overfitting to secret patterns, which could otherwise arise from the high variability in secret lengths in the SecretBench dataset (average length 88.3 characters, standard deviation 342.84).
Figure~\ref{fig:CodeContext} illustrates an example of a secret context.

\subsection{Deduplication Analysis}

Let $C = \{c_1, c_2, \dots, c_n\}$ be the multiset of all $n$ contexts in the dataset. 
We partition $C$ into three mutually exclusive subsets:  the set of exact duplicates ($C_{exact}$), the set of near duplicates ($C_{near}$), and the set of unique contexts ($C_{unique}$).


Specifically, we define what constitutes each subset based on secret contexts.

\subsubsection{Exact Duplicates ($C_{exact}$)}
Exact duplicates are contexts that are identical to one or more other contexts in the dataset. We define $C_{exact}$ multiset as the union of all equivalence classes of identical contexts that contain more than one element.


\emph{Set of representatives ($R_{exact}$).} From the multiset $C_{exact}$, we define a representative set $R_{exact}$. This set is formed by selecting one sample from each distinct group of exact duplicates found in $C_{exact}$. $R_{exact}$ contains one copy of every duplicated context and $R_{exact} \subseteq C_{exact}$.

\subsubsection{Near Duplicates ($C_{near}$)}

Near duplicates are semantically similar contexts that are not identical and differ only in a few aspects.
For instance, the two sample contexts in Listing~\ref{lst:Nearduplicates} are nearly identical, differing only in the values provided during initialization and in the arguments passed to methods.

\definecolor{mygray}{rgb}{0.92, 0.97, 0.99}
\definecolor{myblue}{rgb}{0.13,0.13,1}
\definecolor{mypurple}{rgb}{0.58,0,0.82}
\definecolor{myred}{rgb}{0.7,0,0}
\definecolor{mygreen}{rgb}{0,0.5,0}

\lstset{
  backgroundcolor=\color{mygray},
basicstyle=\ttfamily\footnotesize,
  numbers=left,
  numberstyle=\tiny\color{gray},
  numbersep=5pt,
  xleftmargin=2em,
  frame=single,
  rulecolor=\color{black},
  keywordstyle=\color{myblue}\bfseries,
  stringstyle=\color{myred},
  commentstyle=\color{mygreen},
  identifierstyle=\color{black},
  emph={self,super,def},
  emphstyle=\color{mypurple},
  showstringspaces=false,
  linewidth=0.98\linewidth,
  tabsize=2,
  language=Python,
  breaklines=true,
  breakatwhitespace=true
}

\begin{lstlisting}
TestFramework):
    def __init__(self):
        super().__init__(4, 0)
        self.sporkprivkey = "XXX"

    def run_test(self):
        self.sporkAddress = self.nodes[0].getaccountaddress("")
\end{lstlisting}

\begin{lstlisting}
      super().__init__(6, 5, extra_args=[['-debug=instantsend']] * 6 )
        self.sporkprivkey = "XXX"

    def run_test(self):
        self.sporkAddress = self.nodes[0].getaccountaddress("s")
\end{lstlisting}

\begin{minipage}{\linewidth}
\captionof{lstlisting}{Near-duplicate secrets (ID: 97471 \& 97469). The actual secret values are masked with ``XXX'' to preserve confidentiality.}
\label{lst:Nearduplicates}
\end{minipage}
\\



\subsubsection{Uniques ($C_{unique}$)}
A context is unique if it does not exactly match any other secret context. Uniques are sample contexts in $C_{\neg exact}$ that are neither exact duplicates nor near-duplicates of any other context in the dataset.

 \begin{center}
    $C = C_{unique} \cup C_{near} \cup C_{exact}$
\end{center}

Finally, we construct a deduplicated dataset, $C_{dedup}$ composed of three distinct, non-overlapping parts: all samples from the unique set ($C_{unique}$), all samples from the near-duplicate set ($C_{near}$), and the set of representatives, $R_{exact}$. 
Formally, the complete deduplicated dataset is the union of these three sets:

\begin{center}
$C_{dedup} = C_{unique} \cup C_{near} \cup R_{exact}$
\end{center}


\subsection{Duplication in SecretBench}

We first partition the SecretBench dataset into $C_{exact}$ and $R_{exact}$. From the remaining samples, we then follow the approach of the Near-Duplicate Code Detector \cite{33595913359735} to compute $C_{near}$.

We tokenize the secret context to extract all identifier and literal tokens. 
We construct two types of fingerprints for each context $c_i \in C_{\neg exact}$: a set $T_0$ capturing all unique identifiers and literals, and a multiset $T_1$ reflecting their respective frequencies. Two contexts $i$ and $j$ are classified as duplicates if both the Jaccard similarity between their sets, $J(T_0^i, T_0^j)$, and between their multisets, $J(T_1^i, T_1^j)$, exceed the thresholds $t_0 = 0.8$ and $t_1 = 0.7$, respectively.
All remaining context samples that are neither exact nor near duplicates are classified in $C_{unique}$.

\newpage
We noted a high degree of duplication in the dataset as listed in Table~\ref{tab:duplication_statistics_sb}.
It comprises 97,479 samples, with the 67.523 samples belonging to $C_{exact}$, and 8,472 samples belong to $C_{near}$. Notably, within the $C_{exact}$, only $R_{exact}=6,837$ representative samples exist.
Accordingly, the deduplicated dataset $C_{dedup}$ comprises 36,793 samples, of which 5,290 are true secrets. 


\begin{table}[htpb]
\centering
\caption{Duplications in the SecretBench Dataset.}
\renewcommand{\arraystretch}{1.3}
\label{tab:duplication_statistics_sb}
\begin{tabular}{|l|c|}
\hline
\rowcolor{gray!10}
\textbf{Duplication Types} & \textbf{Count (\%)} \\ \hline
\hline
$C_{unique}$ & 21,484 (22.0\%) \\ 
$C_{exact}$  & 67,523 (69.3\%) \\ 
$C_{near}$   & 8,472 (8.7\%) \\ 
\hline
\textbf{Total} & \textbf{97,479 (100.0\%)} \\
\hline
\end{tabular}
\end{table}

To further investigate the impact of duplication, we assess its effects on the performance and generalization of different models.

\section{Methodology}
\label{sec:Methodology}

%

This section details our dataset construction, experimental scenarios, model architectures, and evaluation metrics.

\subsection{Dataset}\label{sec:DatasetSelection}


Due to the computational complexity of training models on the entire dataset, we filtered SecretBench~\cite{secretbench} to focus on secrets in seven object-oriented programming languages. This selection yields a subset of 23,352 samples, including 3,134 true secrets. 
Table~\ref{tab:duplication_statistics_filtered} summarizes the distribution of all duplication categories in this dataset.

There are 13,779 samples in $C_{exact}$ of which $R_{exact} = 653$ are representative.
Accordingly, $C_{dedup}$ consists of 10,226 samples, of which 1,542 are true secrets.






%

\begin{table}[htpb]
\centering
\caption{Duplications in Filtered SecretBench.}
\renewcommand{\arraystretch}{1.3}
\label{tab:duplication_statistics_filtered}
\begin{tabular}{|l|c|}
\hline
\rowcolor{gray!10}
\textbf{Duplication Types} & \textbf{Count (\%)} \\ \hline
\hline
$C_{unique}$ & 7,372 (31.6\%) \\
$C_{exact}$  & 13,779 (59.0\%) \\
$C_{near}$   & 2,201 (9.4\%) \\
\hline
\textbf{Total} & \textbf{23,352 (100.0\%)} \\
\hline
\end{tabular}
\end{table}

\subsection{Training and Testing}\label{sec:Evaluation}
\begin{figure}[h]
\centering
\scalebox{1}{ 
\begin{tikzpicture}[
    tinyPie/.style = {scale=0.45}, 
    colhead/.style = {font=\small, fill=gray!15, rounded corners=0.1pt,
                      minimum height=0.6cm, minimum width=2.2cm, align=center},
    rowhead/.style = {font=\small, minimum width=1.6cm, align=right},
    every node/.style = {font=\scriptsize}
]

\definecolor{myexact}{RGB}{80, 169, 230} 
\definecolor{mypink}{RGB}{255, 102, 102} 
\definecolor{mynear}{RGB}{255, 204, 102} 
\definecolor{myunique}{RGB}{120, 219, 163} 

\def\xTrain{2.5}
\def\xTest{4.5}  
\def\yMixed{1.2}
\def\yNear{0}
\def\yUnique{-1.2}
\def\yHead{2.5}

\begin{scope}[on background layer]
  \fill[gray!5, rounded corners=2pt] (1.5, -2.0) rectangle (3.5, 2.0);
  \fill[gray!5, rounded corners=2pt] (2.0, -2.0) rectangle (5.5, 2.0);
\end{scope}

\node[colhead, anchor=north] at (\xTrain, \yHead) {Train 80\%};
\node[colhead, anchor=north] at (\xTest, \yHead) {Test 20\%};

\node[rowhead] at (0.5, \yMixed)  {1) Mixed};
\node[rowhead] at (0.5, \yNear)   {2) Near};
\node[rowhead] at (0.5, \yUnique) {3) Unique};

\newcommand{\doughnutPie}[3]{
  \begin{scope}
    \fill[#1] (0,0) -- (90:1cm) arc (90:210:1cm) -- cycle;
    \fill[#2] (0,0) -- (210:1cm) arc (210:330:1cm) -- cycle;
    \fill[#3] (0,0) -- (330:1cm) arc (330:450:1cm) -- cycle;
    \fill[white] (0,0) circle (0.35cm); 
  \end{scope}
}

\newcommand{\uniqueTrainDoughnut}{
  \begin{scope}
    \fill[myexact] (0,0) -- (90:1cm) arc (90:270:1cm) -- cycle;
    \fill[mynear]  (0,0) -- (270:1cm) arc (270:450:1cm) -- cycle;
    \fill[white] (0,0) circle (0.35cm);
  \end{scope}
}

\newcommand{\uniqueTestDoughnut}{
  \fill[myunique] (0,0) circle (1cm);
  \fill[white] (0,0) circle (0.35cm);
}

\begin{scope}[shift={(\xTrain, \yMixed)}, tinyPie]
  \doughnutPie{myexact}{mynear}{myunique}
\end{scope}

\begin{scope}[shift={(\xTest, \yMixed)}, tinyPie, opacity=1]
  \doughnutPie{myexact}{mynear}{myunique}
\end{scope}

\begin{scope}[shift={(\xTrain, \yNear)}, tinyPie]
  \doughnutPie{mypink}{mynear}{myunique}
\end{scope}

\begin{scope}[shift={(\xTest, \yNear)}, tinyPie, opacity=1]
  \doughnutPie{mypink}{mynear}{myunique}
\end{scope}

\begin{scope}[shift={(\xTrain, \yUnique)}, tinyPie]
  \uniqueTrainDoughnut
\end{scope}

\begin{scope}[shift={(\xTest, \yUnique)}, tinyPie, opacity=1]
  \uniqueTestDoughnut
\end{scope}

\begin{scope}[shift={(0.9, -2.5)}]
    \node at (0,0) [anchor=west] {\tikz{\draw[fill=myexact, draw=black, rounded corners=1pt] (0,0) rectangle +(0.25,0.25);}~ $C_{exact}$};
    \node at (1.3,0) [anchor=west] {\tikz{\draw[fill=mypink, draw=black, rounded corners=1pt] (0,0) rectangle +(0.25,0.25);}~ $R_{exact}$};
    \node at (2.6,0) [anchor=west] {\tikz{\draw[fill=mynear, draw=black, rounded corners=1pt] (0,0) rectangle +(0.25,0.25);}~ $C_{near}$};
    \node at (3.9,0) [anchor=west] {\tikz{\draw[fill=myunique, draw=black, rounded corners=1pt] (0,0) rectangle +(0.25,0.25);}~ $C_{unique}$};
\end{scope}

    \end{tikzpicture}}
\caption{Three Main Train/Test Scenarios.}
\label{Fig:scenario_summary}
\end{figure}
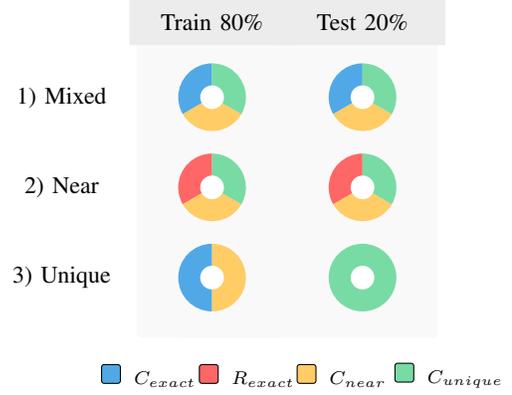

To quantify the impact of data duplication on model performance, we design three distinct evaluation scenarios using our dataset. For each scenario, we establish a fixed 5-fold cross-validation strategy. We create these 5 folds (each holding out 20\% of the data for testing) in advance. This ensures that all models and experiments use the same data splits, allowing for a consistent and fair comparison. Within each fold, 10\% of the training data is further allocated as a validation set. Figure \ref{Fig:scenario_summary} summarizes the specific data composition used for training and testing in each of these three scenarios.

\subsubsection{Mixed (Baseline on Duplicated Data)}
This scenario measures model performance in the presence of all duplicates, simulating a naive evaluation on unpreprocessed data. We used the complete subset of 23,352 records for both training and testing.

\subsubsection{Near Duplicate (Baseline on Deduplicated Data)}
This scenario establishes the model’s baseline performance without exact duplications. We use the deduplicated dataset $C_{dedup}$, defined as $C_{dedup} = C_{unique} \cup C_{near} \cup R_{exact}$ for training and testing.
Since this scenario applies cross-validation, both the training and test splits within each fold will contain samples from the $C_{near}$ set. This design specifically allows us to isolate and measure the impact of near-duplicates on model performance, as the model is trained and evaluated on datasets containing different, but semantically similar, instances.

\subsubsection{Unique (Training on Duplicates, Testing on Unique Data)}
This scenario is designed to directly measure the model's ability to generalize from redundant data to entirely novel samples. We construct the training dataset by combining all samples from the $C_{exact}$ and $C_{near}$. In case any samples from the unique set $C_{unique}$ are left after selecting the test set, they are added to the train set. The model is then evaluated exclusively on the held-out unique subset, $C_{unique}$.

By comparing performance across these three scenarios, we can isolate and quantify the influence of exact and near duplicates on model evaluation.

\subsection{Models}
We adopt four models with the Differential Evolution (DE) algorithm to optimize hyperparameters and fine tune models iteratively. 

\emph{Random Forest (RF).} 
RF is an ensemble learning algorithm that constructs multiple decision trees during training to achieve stable and accurate predictions. It then aggregates the outputs of these trees using a voting mechanism for prediction. RF has proven to be effective in complex defect detection \cite{10.1145/3689236.3695374}. 
We employ the \textit{RandomForestClassifier} from the \textit{scikit-learn} library to construct our model.
We optimize the key hyperparameters using DE. The hyperparameters tuned include the number of estimators, maximum depth, minimum samples split, minimum samples leaf, and maximum features. Each candidate hyperparameter set is assessed using the \textit{rf\_tuning} function, which fits the model to the training data and calculates the F1-score, serving as the fitness function to identify the best-performing hyperparameters.

\emph{Long Short-Term Memory (LSTM).} LSTM networks are particularly well-suited for deep learning analysis of sequential data, making them ideal for identifying code patterns \cite{hochreiter1997long}. Their architecture is designed to capture long-term dependencies, which are crucial for understanding the sequences in code snippets where secrets might be embedded. We utilize the LSTM model from the \textit{TensorFlow Keras} library. We propose an enhanced LSTM network architecture for secret detection. This architecture increases the depth of the LSTM network by employing a stack of two LSTM layers, accompanied by dropout layers for regularization. 

The architecture dynamically adapts to input based on \textit{vocab\_size}. For text with \textit{vocab\_size} \texttt{>} 1, the input passes through an embedding layer, converting words into dense vectors. This is followed by two LSTM layers, each with dropout to prevent overfitting. This path captures complex sequential patterns. If \textit{vocab\_size} \texttt{<=} 1, a dense layer with ReLU activation processes numerical data. Both paths lead to a final dense layer with a sigmoid activation, outputting a binary classification, which serves as the classifier for detecting secrets. This setup effectively handles different feature types by adjusting its processing strategy.  We experimented with different units for the first LSTM layer (128, 192, and 256) and varied the number of epochs (5, 10, and 20) to determine the optimal configuration.

\emph{GraphCodeBERT-based Models.} We leverage GraphCodeBERT \cite{GraphCodeBERT}, a transformer-based model designed to capture both the context and inherent structure of source code. Unlike approaches that rely on syntactic-level structures such as Abstract Syntax Trees (ASTs), GraphCodeBERT incorporates semantic-level information by modeling data flow as a graph, where nodes correspond to variables and edges indicate the “where-the-value-comes-from” relationships. Pre-trained on the CodeSearchNet dataset \cite{CodeSearchNet}, which contains 2.3 million functions across six programming languages (Go, Java, JavaScript, PHP, Python, and Ruby), GraphCodeBERT effectively encodes code semantics for downstream tasks.

To adapt the model for our dataset, inputs are tokenized using the \textit{AutoTokenizer} for \textit{microsoft/GraphCodeBERT-base}, producing subword token IDs and attention masks. Padding is applied dynamically within each batch, and attention masks ensure the model attends only to meaningful tokens. These token IDs and masks are wrapped in a TensorDataset and loaded via a DataLoader with random sampling for training and evaluation.

We consider two variants of GraphCodeBERT for classification:

\emph{Full Fine-Tuning (GraphCodeBERT).}In this standard approach, all model parameters are unfrozen and updated during training, allowing the model to fully adapt to the target dataset.

\emph{Feature Extraction with MLP (GraphCodeBERT-MLP).} Here, the main GraphCodeBERT parameters are frozen, and the model serves as a static feature extractor. Token embeddings are mean-pooled to generate a single 768-dimensional vector per input, which is then passed to a lightweight Multi-Layer Perceptron (MLP) classifier. The MLP consists of three hidden linear layers with ReLU activations and Dropout, followed by a softmax output layer. This strategy reduces training time and memory usage while providing strong regularization.

Using 5 consistent training and testing splits across all experiments, we employ four learning approaches combined with three scenarios and a feature. This results in
\( 1 feature \times 4 learners \times 3 scenarios \times 5 ways = 60\) models for secret detection.
All reported metrics for these models are averaged over 5 training and testing splits. Note that all 5 splits are consistent when applying models using learners in each scenario. 

We ran all experiments on a MacBook Air equipped with an M3 chip and 24 GB of RAM.

\subsection{Performance Metrics}

The prediction results are grouped into four categories, as summarized in Table~\ref{tab:ConfusionMatrix-table}.
\begin{table}[htbp]
\centering
\caption{Confusion Matrix.}
\label{tab:ConfusionMatrix-table}
\setlength{\tabcolsep}{4pt}
\renewcommand{\arraystretch}{1.3} 
\begin{tabular}{lp{0.24\columnwidth}p{0.24\columnwidth}p{0.24\columnwidth}} 
 & \multicolumn{1}{c}{} & \multicolumn{2}{c}{Prediction} \\ \cline{3-4} 
 & \multicolumn{1}{l|}{} & \multicolumn{1}{l|}{\textbf{Secrets}} & \multicolumn{1}{l|}{\textbf{Non-secrets}} \\ \cline{2-4} 
\multicolumn{1}{l|}{\multirow{2}{*}{Actual}}  & \multicolumn{1}{l|}{\textbf{Secrets}} & \multicolumn{1}{l|}{True Positive (TP)} & \multicolumn{1}{l|}{False Negative (FN)} \\ \cline{2-4} 
\multicolumn{1}{l|}{} & \multicolumn{1}{l|}{\textbf{Non-secrets}} & \multicolumn{1}{l|}{False Positive (FP)} & \multicolumn{1}{l|}{True Negative (TN)} \\ \cline{2-4} 
\end{tabular}%
\end{table}

The evaluation metrics in this study include: \textit{Precision}, which measures the accuracy of positive predictions (Equation.\ref{eq:precision}); \textit{Recall}, indicating the probability of detection (Equation.\ref{eq:recall}); \textit{F1-score}, the harmonic mean of precision and recall (Equation.\ref{eq:f1score}); 
While F1-score is a well-accepted evaluation metric, it does not consider all the quadrants of a confusion matrix \ref{tab:ConfusionMatrix-table}. Consequently, this can result in inflated results when the dataset is imbalanced. To address this, we also report the Matthews correlation coefficient (MCC) (Equation.\ref{eq:mcc}) \cite{MCC}.

\begin{small}
\begin{equation}
    \text{Precision} = \frac{TP}{TP + FP} \label{eq:precision}
\end{equation}
\end{small}

\begin{small}
\begin{equation}
    \text{Recall} = \frac{TP}{TP + FN} \label{eq:recall}
\end{equation}
\end{small}

\begin{small}
\begin{equation}
    \text{F1-score} = \frac{2 \times \text{Recall} \times \text{Precision}}{\text{Recall} + \text{Precision}} \label{eq:f1score}
\end{equation}
\end{small}

\begin{small}
\begin{equation}
    \text{MCC} = \frac{TP \times TN - FP \times FN}{\sqrt{(TP+FP)(TP+FN)(TN+FP)(TN+FN)}} 
    \label{eq:mcc}
\end{equation}
\end{small}

\section{RESULTS}
\label{sec:EXPERIMENTRESULTS}
In this section, we present the results of three scenarios in addressing the following research question:

\begin{tcolorbox}[RQsboxstyle]
    \textbf{RQ:} To what extent does code duplication impact the performance and generalization of machine learning models for secret detection?
\end{tcolorbox}



\subsection{Mixed Scenario} 
As presented in Table~\ref{tab:Mixed}, all models demonstrated notably high performance metrics in this scenario, where the test set includes exact and near duplicates of training samples. GraphCodeBERT significantly outperformed (p-value $<$ 0,05) other models, achieving near-perfect results with both F1-score and MCC reaching 0.97. 
We observed no statistically significant difference between the performance of the LSTM and GraphCodeBERT-MLP models (with a $p$-value of $0.12$).
While the Random Forest followed closely with an MCC of 0.91, GraphCodeBERT-MLP and LSTM performed statistically significantly better than the Random Forest.

\begin{table}[htpb]
\centering
\caption{Mixed Scenario Results.}
\renewcommand{\arraystretch}{1.3}
\label{tab:Mixed}
\begin{tabular}{|l|c|c|c|c|}
\hline
\rowcolor{gray!10}
\textbf{Classifier} & \textbf{Precision} & \textbf{Recall} & \textbf{F1-Score} & \textbf{MCC} \\ \hline
\hline
GraphCodeBERT & 0.97 & 0.98 & 0.97 & 0.97 \\ \hline
GraphCodeBERT-MLP & 0.92 & 0.93 & 0.93 & 0.92 \\ \hline
Random Forest & 0.96 & 0.86 & 0.91 & 0.89 \\ \hline
LSTM & 0.92 & 0.94 & 0.93 & 0.92 \\
\hline
\end{tabular}
\end{table}

\begin{table}[htpb]
\centering
\caption{Near Duplicate Scenario Results.}
\renewcommand{\arraystretch}{1.3}
\label{tab:scenario_fully_unbiased}
\begin{tabular}{|l|c|c|c|c|}
\hline
\rowcolor{gray!10}
\textbf{Classifier} & \textbf{Precision} & \textbf{Recall} & \textbf{F1-Score} & \textbf{MCC} \\ \hline
\hline
GraphCodeBERT & 0.91 & 0.96 & 0.93 & 0.92 \\ \hline
GraphCodeBERT-MLP & 0.85 & 0.83 & 0.84 & 0.81 \\ \hline
Random Forest & 0.90 & 0.71 & 0.79 & 0.77 \\ \hline
LSTM & 0.84 & 0.84 & 0.84 & 0.82 \\
\hline
\end{tabular}
\end{table}

\subsection{Near Duplicate Scenario}
The results presented in Table \ref{tab:scenario_fully_unbiased} show that, similar to the Mixed scenario, GraphCodeBERT obtained a statistically significantly higher MCC than all three other models, with an MCC of 0.92 (paired t-test, $p < 0.05$).
The comparison between LSTM and GraphCodeBERT-MLP revealed no statistically significant difference ($p > 0.89$) with MCC values of 0.82 and 0.81, respectively. While these models showed comparable performance, both were significantly better than RF, with the lowest MCC of 0.77 ($p < 0.05$).


\begin{table}[htpb]
\centering
\caption{Unique Scenario Results.}
\renewcommand{\arraystretch}{1.3}
\label{tab:scenario_unbiased}
\begin{tabular}{|l|c|c|c|c|}
\hline
\rowcolor{gray!10}
\textbf{Classifier} & \textbf{Precision} & \textbf{Recall} & \textbf{F1-Score} & \textbf{MCC} \\ \hline
\hline
GraphCodeBERT & 0.91 & 0.92 & 0.91 & 0.90 \\ \hline
GraphCodeBERT-MLP & 0.82 & 0.80 & 0.81 & 0.78 \\ \hline
Random Forest & 0.85 & 0.56 & 0.67 & 0.65 \\ \hline
LSTM & 0.82 & 0.79 & 0.80 & 0.77 \\ 
\hline
\end{tabular}
\end{table}

\subsection{Unique Scenario} 

As detailed in Table~\ref{tab:scenario_unbiased}, GraphCodeBERT performed significantly better than all other models with an MCC of 0.90.
Additionally, GraphCodeBERT-MLP and LSTM showed comparable results, with no statistically significant difference observed; their MCC values were 0.78 and 0.77, respectively.
In contrast, RF struggled to generalize with an MCC of 0.77.

\subsection{
Scenario Comparison
}

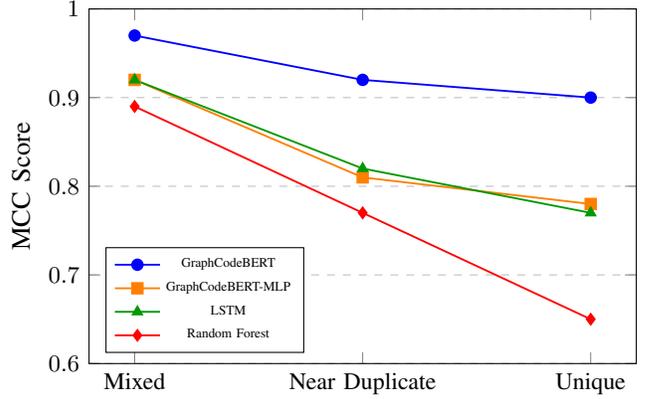
\begin{figure}[htbp]
    \centering
    \begin{tikzpicture}
    \begin{axis}[
        width=\linewidth,    
        height=6.3cm,             
        xlabel={},
        ylabel={MCC Score},
        ymin=0.6, ymax=1.0,     
        xtick={1,2,3},
        xticklabels={Mixed, Near Duplicate, Unique},
        x tick label style={font=\small},
        y tick label style={font=\small},
        legend pos=south west,  
        legend style={font=\footnotesize},
        ymajorgrids=true,
        grid style=dashed,
        title={},
        legend style={font=\tiny}
    ]

    \addplot[
        color=blue,
        mark=*,
        line width=0.7pt
        ]
        coordinates {
        (1,0.97)(2,0.92)(3,0.90)
        };
        \addlegendentry{ GraphCodeBERT}

    \addplot[
        color=orange,
        mark=square*,
        line width=0.7pt
        ]
        coordinates {
        (1,0.92)(2,0.81)(3,0.78)
        };
        \addlegendentry{GraphCodeBERT-MLP}

    \addplot[
        color=green!60!black,
        mark=triangle*,
        line width=0.7pt
        ]
        coordinates {
        (1,0.92)(2,0.82)(3,0.77)
        };
        \addlegendentry{LSTM}

    \addplot[
        color=red,
        mark=diamond*,
        line width=0.7pt
        ]
        coordinates {
        (1,0.89)(2,0.77)(3,0.65)
        };
        \addlegendentry{Random Forest}

    \end{axis}
    \end{tikzpicture}
    \caption{MCC performance across test scenarios.}
    \label{fig:mcc_degradation}
\end{figure}

Figure~\ref{fig:mcc_degradation} illustrates the MCC scores across the three scenarios.
We observe a clear downward trend in performance across the scenarios: 
the highest scores occur in the Mixed scenario, followed by a decrease in the Near Duplicate scenario, with the lowest performance in the Unique scenario. This pattern indicates that removing data leakage leads to a drop in predictive capability for all models.
GraphCodeBERT consistently achieves the highest MCC scores. 
Its MCC degradation of only 7.22\% highlights its robustness to duplication.
The fully fine-tuned GraphCodeBERT demonstrates greater resilience to duplication than its frozen-encoder variant with an MLP head. However, training the GraphCodeBERT-MLP model is substantially faster (approximately 1 hour versus 8 hours).
Interestingly, the additional effort required for GraphCodeBERT-MLP may not be justified, as LSTM achieves comparable performance with only minimal differences.
RF is the lowest-performing model. It demonstrates a good performance in the Mixed scenario, but drops sharply by 26.97\% in the Unique scenario, showing its failure to generalize beyond duplicated contexts.

\begin{tcolorbox}[blueboxstyle]
GraphCodeBERT demonstrates superior performance among existing models, maintaining an MCC score of minimum 90\% in all scenarios. In contrast, other models are more vulnerable to duplication, with RF exhibiting the weakest resilience, dropping from 89\% to 65\%.
\end{tcolorbox}

\newpage
\section{THREATS TO VALIDITY}
\label{sec:ThreatsToValidity}

\emph{Internal validity.}
Our design choices, such as the use of a 200-character context window (excluding the secret string), may influence results; alternative configurations were not systematically tested.
Outlier secret lengths could bias models toward moderate-length secrets and limit generalization. 
Model variability and differences in configurations may affect reproducibility.
Furthermore, duplicates in the SecretBench inflated model performance by allowing memorization; removing these enabled a more accurate assessment of real-world effectiveness.

\emph{External validity.}
Our use of the SecretBench dataset may limit the generalizability of our findings, as it is constructed from secrets identified by TruffleHog and GitLeaks, potentially omitting certain types of secrets. Because SecretBench is not publicly available and access is restricted, models are unlikely to have encountered it during pretraining, which reduces the risk of prior exposure but limits reproducibility. While we document our deduplication process and release code, replication requires dataset access via the original authors.

\emph{Construct validity.}
We used the SecretBench dataset as our ground truth; however, potential mislabelings from human annotation may pose a threat to construct validity.

\section{Conclusion}
\label{sec:Conclusion}

We demonstrate that duplications in the SecretBench dataset can significantly inflate the reported performance of AI-based secret detectors. Models are not inherently robust to such duplications, and their sensitivity varies across different model families, sometimes resulting in up to a 26\% higher MCC. This suggests that the exceptional performance reported in the literature may reflect memorization rather than true learning.

We call for the reevaluation of secret detection models to address data leakage and ensure that models are reliable and applicable in real-world settings.

%

\section*{Data Availability}

The dataset and the full replication package are publicly available: https://zenodo.org/records/18164088

\balance
\bibliographystyle{IEEEtran}
\bibliography{references}

\end{document}